# Generalized Lorenz dominance orders

Leo Egghe[1] and Ronald Rousseau[2,3] *

[1] Hasselt University, 3500 Hasselt, Belgium

leo.egghe@uhasselt.be

ORCID: 0000-0001-8419-2932

[2] KU Leuven, MSI, Facultair Onderzoekscentrum ECOOM,

Naamsestraat 61, 3000 Leuven, Belgium

ronald.rousseau@kuleuven.be   &

[3] University of Antwerp, Faculty of Social Sciences,

Middelheimlaan 1, 2020 Antwerp, Belgium

ronald.rousseau@uantwerpen.be

ORCID: 0000-0002-3252-2538

ABSTRACT

We extend the discrete majorization theory by working with non-normalized Lorenz curves. Then we prove two generalizations of Muirhead's theorem. They not only use elementary transfers but also local increases. Together these operations are described as elementary impact increases. The first generalization shows that if an array X is dominated, in the generalized sense, by an array Y then Y can be derived from X by a finite number of elementary impact increases and this in such a way that each step transforms an array into a new one which is strictly larger in the generalized majorization sense. The other one shows that if the dominating array, Y, is ordered decreasingly then elementary impact increases starting from the dominated array, X, lead to the dominating one. Here each step transforms an array to a new one for which the decreasingly ordered version dominates the previous one and is dominated by Y.

* Corresponding author

## 1. Introduction

In this work, situated within the framework of dominance studies and related investigations, we make further progress in our investigations on these subjects. We formulate and provide proofs of two generalized Muirhead theorems (explained further on).

Many distributions studied in informetrics, such as authors and their publications (Lotka, 1926), or topics and journals dealing with them (Bradford, 1934) can be described by power law relations (Egghe, 2005) or similar skewed distributions. One common characteristic of these distributions is the high concentration of items among a few sources. As such the study of concentration or inequality, with its social implications, is one of the main topics studied in the science of science (Rousseau et al., 2018, Section 9.5).

Equality or evenness, the opposite of inequality or concentration, plays a key role in studies of interdisciplinarity or, more generally, diversity studies (Wagner et al., 2011; Rousseau et al., 2019). Of course, concentration and diversity are also essential notions in other fields such as economics and ecology, and have widespread implications.

We recall here that Muirhead's theorem, also referred to as Muirhead's inequality (Muirhead, 1903), and the Lorenz curve (Lorenz, 1905), both formally defined below, form the basis of welfare economics (Kleiber, 2008). Moreover, virtually all generalizations of the arithmetic-geometric mean inequality follow from Muirhead's theorem. In his (econometric) review article, Kleiber (2008), wrote that the Lorenz curve has considerable further potential, i.e., further than considered in his review, in theoretical and applied economics.

For this reason, we expect that generalizations of Muirhead's Theorem will lead to new opportunities for studies in the science of science and beyond.

## 2. Notations and some preliminaries

Let $N > 0$ be a natural number, $\mathbb{R}^+$ denotes the non-negative real numbers, and let $X = (x_1, \cdots, x_N)$ be an $N$-array, with $x_j \in \mathbb{R}^+, j = 1, \cdots, N$. The set of all these $N$-arrays is denoted as $(\mathbb{R}^+)^N$.

Definition. The classical majorization order

If $X$ and $Y$ are $N$-arrays then $X$ is majorized by $Y$ (Y majorizes X), denoted as $X -<_L Y$ if

$$\sum_{j=1}^{k} x_{[j]} \leq \sum_{j=1}^{k} y_{[j]} \quad \text{for } k = 1, \cdots, N-1 \qquad (1)$$

and $\qquad \sum_{j=1}^{N} x_j = \sum_{j=1}^{N} y_j,$

where $(x_{[j]})$, j=1,…,N and $(y_{[j]})$, j=1,…,N, denote the components of X and Y when ranked in decreasing order (this is the same notation as used in (Marshall et al., 2011)).

This terminology originates from (Hardy et al., 1934, 1952). The index L in $X -<_L Y$ refers to the fact that this order relation corresponds to the order relation derived from the corresponding Lorenz curves. For further use, we recall that the Lorenz curve of array $X$ is the curve in the plane obtained by the line segments connecting the origin (0,0) to the points $\left(\frac{k}{N}, \frac{\sum_{j=1}^{k} x_{[j]}}{\sum_{j=1}^{N} x_{[j]}}\right)$, $k = 1, \cdots, N$. For $k = N$, one reaches the endpoint (1,1). One sees that $X$ is majorized by $Y$, $(X -<_L Y)$, iff the Lorenz curve of $Y$ is situated above the Lorenz curve of $X$.

It is well-known, see e.g. (Marshall et al., 2011, p.14) that $X -<_L Y$ is equivalent with each of the following statements:

(A) $\sum_i \varphi(x_i) \le \sum_i \varphi(y_i)$ for all continuous, convex functions $\varphi: \mathbb{R} \to \mathbb{R}$.

(B) Y can be obtained from X by a finite number of elementary transfers (Muirhead, 1903)

Here an elementary transfer is a transformation from $(x_1, \cdots, x_N)$, , $(x_1, \cdots, x_N)$ ranked in decreasing order, into $(x_1, \ldots, x_i + h, \ldots, x_j - h, \ldots, x_N)$ where $0 < h \le x_j$.

Definition. A discrete concentration function

A function m from $(\mathbb{R}^+)^N$ to the positive real numbers is a concentration function if it is an order morphism from $(\mathbb{R}^+)^N, -<_L$ to the positive real numbers. This means that X-$<_L$ Y implies that m(X) $\le$ m(Y), with equality only if X and Y have the same Lorenz curve.

## 3. Generalized majorization and Muirhead-type theorems

Definition. Elementary impact increase

An elementary impact increase, in short EII, is a function that maps the array $X = (x_1, \cdots, x_N)$ to $EII(X) = (x_1, \cdots, x_i + a, \ldots, x_j - b, \ldots, x_N)$ with a > 0, and either b=a > 0 (and then $x_j \ge$ b=a > 0) or b=0, $i \in \{1, \ldots, N\}, j \in \{i + 1, \ldots, N\}$.

We see that if b > 0 then an elementary impact increase is a transfer in the sense of Muirhead-Dalton (Marshall et al., 2011).

Definition. Generalized majorization

Given the natural number N > 0, we define the generalized majorization relation on $(\mathbb{R}^+)^N$, denoted as -<, by X -< Y iff

$$I_X(k) = \sum_{j=1}^{k} x_j \leq I_X(k) = \sum_{j=1}^{k} y_j \ \text{ for } k = 1, \cdots, N \tag{2}$$

Clearly X=Y $\Leftrightarrow I_X = I_Y$ and if X≠Y then there exist at least on j = 1, …, N, such that $I_X(j) < I_Y(j)$.

Remark that contrary to the classical majorization as applied to Lorenz curves, we do not require in (2) that X (and Y) are ordered in size and neither have sums of their arrays to be equal. This means that such arrays can be used to represent a timeline (over a fixed period).

We see that the relation -< is by definition reflexive, antisymmetric, and transitive, hence a partial order (Roberts, 1979, p.15).

Proposition. $\forall X \in (\mathbb{R}^+)^N$: X -< EII(X), strictly.

Proof. Consider A = EII(X) = $(x_1, \cdots, x_i + a, \ldots, x_j - b, \ldots, x_N)$. For simplicity, we denote the components of A as $(a_j)_j$. Clearly X ≠ A. Then

$$\text{if } i > 1: \forall k, k < i : \sum_{l=1}^{k} a_l = \sum_{l=1}^{k} x_l$$

$$\text{if } j > i: \forall k, i \leq k < j - 1: \sum_{l=1}^{k} a_l = \sum_{l=1}^{k} x_l + a > \sum_{l=1}^{k} x_l$$

$$\forall k, j \leq k \leq N: \sum_{l=1}^{k} a_l = \sum_{l=1}^{k} x_l + (a - b) \geq \sum_{l=1}^{k} x_l$$

This shows that X -< EII(X), strictly.□

A simple example. Let X = $(\sqrt{2}, 7, 0, \pi)$. Then N =4; taking i=1, j= 4, $a = \pi$ and $b_1 = 0$, $b_2 = \pi$, and applying EII on X leads to $(\sqrt{2} + \pi, 7, 0, \pi)$, for $b_1$ and $(\sqrt{2} + \pi, 7, 0, 0)$, for $b_2$. In both cases the transformed 4-tuple is strictly larger than the original in the -< order.

Next, we formulate a Muirhead (1903)-type theorem. It is related to, but different from, Theorem A.5 (v) in (Marshall et al., 2011).

Theorem A:  Muirhead's theorem for the generalized majorization relation.

For the $N$-arrays X and Y in $(\mathbb{R}^+)^N$, X≠Y, we have:

(i) X -< Y

$\Leftrightarrow$

(ii) Y can be derived from X through a finite number of elementary impact increases

Proof.

(ii) $\Rightarrow$ (i) As X -< EII(X) and as the relation -< is transitive this implication is trivial.

(i) $\Rightarrow$ (ii). Let X ≠ Y, and

$$X = (x_1, \cdots, x_N) \text{ -< } Y = (y_1, \cdots, y_N)$$

Case A. If for all $i \in \{1, \ldots, N\}$, $x_i \leq y_i$, with, necessarily, at least one index i such that $x_i < y_i$, then we consider the smallest index i, denoted as $i_0$, for which this strict inequality happens. Then we apply the elementary impact increase for which a = $y_i$ - $x_i$ on position $i_0$ and with b=0. This yields an array $X_1$. Clearly X -< $X_1$ -< Y. The first inequality is strict, while the second one becomes an equality if $i_0$ is the only index for which $x_i < y_i$. Otherwise, the inequality $X_1$ -< Y is also strict. The $i_0$ first components of arrays $X_1$ and Y are certainly equal. Now we can apply the same procedure on $X_1$. As the length of the arrays (N) is finite we see that X becomes equal to Y in a finite number of steps and such that in each step the new array is situated between the previous one and Y (in the generalized majorization order), finally becoming equal to Y. Note that equality may already happen before we reach index N.

Case B. If the relation $x_i \leq y_i$, for all $i \in \{1, \ldots, N\}$, does not hold, we put j equal to the first index such that $x_j > y_j$. Let c = $x_j$-$y_j$. As X -< Y, there must be at least one index strictly smaller than j such that the component of X is strictly smaller than the corresponding component of Y. We denote the set of indices where this happens by M. We choose the first number of indices $i_1 < \cdots < i_k$ in M such that

$$\sum_{l=1}^{k-1}(y_{i_l} - x_{i_l}) < c \text{ and } \sum_{l=1}^{k}(y_{i_l} - x_{i_l}) \geq c$$

If k = 1 then only the second inequality matters and we put the first sum equal to zero. We note that $x_s$ = $y_s$ if s ≤ $i_k$ and s does not belong to the set M. Then we apply the following k elementary impact increases: for $l$ =1, …, k-1: a = $y_{i_l} - x_{i_l}$ = b (if k = 1, then this part is not performed) and for the case $l$ = k: a = $c - \sum_{l=1}^{k-1}(y_{i_l} - x_{i_l})$ = b > 0 (a is applied on index $i_l$ and b on index j).

After $l$ steps, $l = 1, \ldots,$ k, we have a new array $X_{i_l} = (z_1, \ldots, z_{i_l}, x_{i_l+1}, \ldots, x_{j-1}, z_j, x_{j+1}, \ldots, x_N)$. The first $i_l$ coordinates of $X_{i_l}$ are of the form $z_s = x_s + \delta_s$, with $\delta_s \geq 0$ (equal to zero is possible), with $\sum_{s=1}^{i_l} \delta_s = c' \leq c$ and $z_j = x_j - c'$. We see that $y_j \leq z_j < x_j$. When $l = k$, then $y_j = z_j$. At each step, $i_l > 1$, $X_{i_{l-1}} -< X_{i_l} -<$ Y. The first inequality (-<) follows from the observation that (ii) implies (i). Now we prove the second one. We first note that the first $i_{l-1}$ coordinates of $X_{i_l}$ and $X_{i_{l-1}}$ stay invariant until we reach step k. Now we check if $X_{i_l} -<$ Y $(i_l < i_k)$:

$$\sum_{s=1}^{i_l}(X_{i_l})_s = \sum_{s=1}^{i_l} z_s = \sum_{s=1}^{i_l}(x_s + \delta_s) = \sum_{s=1}^{i_l} y_s$$

Now, for $i_l < m < j$, we have, by the definition of the index j, that $\sum_{s=1}^{m}(X_{i_l})_s = \sum_{s=1}^{i_l}(X_{i_l})_s + \sum_{s=i_l+1}^{m}(X_{i_l})_s = \sum_{s=1}^{i_l} y_s + \sum_{s=i_l+1}^{m} x_s \leq \sum_{s=1}^{m} y_s$;

Further, we have: $\sum_{s=1}^{j}(X_{i_l})_s = \sum_{s=1}^{i_l}(x_s + \delta_s) + \sum_{s=i_l+1}^{j-1} x_s + z_j = \sum_{s=1}^{j-1} x_s + c' + x_j - c' \leq \sum_{s=1}^{j} y_s$.

Finally, if $j < m \leq$ N: $\sum_{s=1}^{m}(X_{i_l})_s = \sum_{s=1}^{j}(X_{i_l})_s + \sum_{s=j+1}^{m}(X_{i_l})_s = \sum_{s=1}^{j} x_s + \sum_{s=j+1}^{m} x_s = \sum_{s=1}^{m} x_s \leq \sum_{s=1}^{m} y_s$.

Next, we also show that $X_{i_k} -<$ Y

$$\sum_{s=1}^{i_k}(X_{i_k})_s = \sum_{s=1}^{i_k} z_s = \sum_{s=1}^{i_k}(x_s + \delta_s) = \sum_{s=1}^{i_k} y_s$$

Now, for $i_k < m < j$, we have that $\sum_{s=1}^{m}(X_{i_k})_s = \sum_{s=1}^{i_k}(X_{i_k})_s + \sum_{s=i_k+1}^{m}(X_{i_k})_s = \sum_{s=1}^{i_k} y_s + \sum_{s=i_k+1}^{m} x_s \leq \sum_{s=1}^{m} y_s$;

Further, we have: $\sum_{s=1}^{j}(X_{i_k})_s = \sum_{s=1}^{i_k}(x_s + \delta_s) + (\sum_{s=i_k+1}^{j-1} x_s) + z_j = (\sum_{s=1}^{j-1} x_s) + c + x_j - c \leq \sum_{s=1}^{j} y_s$

Finally, if $j < m \leq$ N: $\sum_{s=1}^{m}(X_{i_k})_s = \sum_{s=1}^{j}(X_{i_k})_s + \sum_{s=j+1}^{m}(X_{i_k})_s = \sum_{s=1}^{j} x_s + \sum_{s=j+1}^{m} x_s = \sum_{s=1}^{m} x_s \leq \sum_{s=1}^{m} y_s$.

Hence, for all $l = 1, \ldots,$ k$_1$: $X -< X_{i_l} -< X_{i_k} -<$ Y.

Now we have that the first j components of the modified array X, denoted as $X_{i_k}$, are all smaller than or equal to the corresponding components of Y and $X_{i_k} -<$ Y. This means that we are either in case A or in case B with a new index j, denoted as j$_0$, with j$_0$ > j, such that $x_{j_0} > y_{j_0}$. In either case, we can apply the same procedure and end in a finite number of steps.

Corollary. The original Muirhead theorem (Muirhead, 1903; Hardy et al., 1934)

For the N-arrays X and Y in $(\mathbb{R}^+)^N$, X≠Y, with $\sum_{i=1}^{N} x_i = \sum_{i=1}^{N} y_i$ we have:

(i) $X -<_L Y$

⇔

(ii) Y can be derived from X through a finite number of elementary transfers, i.e., EIIs with a=b >0

Proof. We have to show that if $\sum_{i=1}^{N} x_i = \sum_{i=1}^{N} y_i$ the case b=0 in an EII never happens.  But if X-< Y and an EII with b=0 is applied, then the total sum of all components of X would be larger than the total sum of the Y-components. As applying an EII never decreases the total sum of all components, all EIIs must be elementary transfers, i.e., EIIs with a=b > 0.□

An application. When comparing two situations over time then the recent past is usually more important than the distant past. So, in the array $X = (x_1, \cdots, x_N)$ and using years as the time unit, $x_1$ refers to last year's item of importance, i.e., number of publications, number of citations received, amount of grants obtained (in monetary units),  while $x_N$ refers to the same item N years ago. Now when is X "better" than Y over this N year period? If for each j =1, …, N, $y_j > x_j$ this is certainly the case,  but one may also agree that this is asking too much. Then the relation X -< Y provides an acceptable answer. Indeed, Theorem A show that when X-< Y then Y can be derived from X through a finite number of elementary increases in the item of importance (trade, publications, citations, grants).

As preparation for another Muirhead-type theorem (Muirhead, 1903) about generalized majorization, we first prove two simple lemmas. If $X \in (\mathbb{R}^+)^N$ , we denote by π(X) the decreasingly ranked version of array X, and by ξ(X) the increasingly ranked X.

Lemma 1. $\forall X \in (\mathbb{R}^+)^N$: ξ(X) -< X -< π(X)

Proof. With ξ(X) = $(c_1, …, c_N)$ and π(X) = $(d_1, …, d_N)$, this result is immediate as $\forall i = 1, …, N: \sum_{j=1}^{i} c_j \leq \sum_{j=1}^{i} x_j \leq \sum_{j=1}^{i} d_j$ .

Lemma 2. For X, Y $\in (\mathbb{R}^+)^N$ , $N$ strictly positive, X ≤ Y, with Y  decreasing, implies π(X) ≤ Y.

Proof. By induction on N

a) Let N =2 and let X = $(x_1, x_2)$ and Y=$(y_1, y_2)$. Then we know that $x_1 \leq y_1$ and $x_2 \leq y_2$. If $x_1 \geq x_2$ then $\pi(X) = (x_1, x_2)$ -< Y. If $x_1 < x_2$ then if $x_1 < x_2 \leq y_2 \leq y_1$, from which it follows that $\pi(X) \leq Y$.

b) Assume that we know Lemma 2 for arrays in $(\mathbb{R}^+)^N$, then we prove the lemma for arrays in $(\mathbb{R}^+)^{N+1}$.

Let X = $(x_1, ..., x_{N+1})$ and Y = $(y_1, ..., y_{N+1})$ and denote $(x_1, ..., x_N)$ by $X_N$ and $(y_1, ..., y_N)$ by $Y_N$. We recall that $Y_N$ is ranked in decreasing order. Then $X_N \leq Y_N$ and by the induction hypothesis we know that $\pi(X_N) = (d_1, ..., d_N) \leq Y_N$. Consider now $x_{N+1}$. We assume that in $\pi(X)$ $x_{N+1}$ is placed on the j-th place, where j can in principle be any index between 1 and N+1. Then $\pi(X) = (d_1, ..., d_{j-1}, x_{N+1}, d_j, ..., d_N)$. Now $\pi(X) \leq Y$, because $d_1 \leq y_1$ ; ...; $d_{j-1} \leq y_{j-1}$ ; $x_{N+1} \leq y_{N+1} \leq y_j$ ; $d_j \leq x_{N+1} \leq y_{N+1} \leq y_{j+1}$ ; ... ; $d_N \leq x_{N+1} \leq y_{N+1}$. This shows that for any N > 0 we have: X, Y $\in (\mathbb{R}^+)^N$ and X $\leq$ Y, with Y decreasing, implies $\pi(X) \leq Y$.

Remark. This lemma does not hold for $\leq$ replaced by -<. Indeed, take X = (1,3) and Y=(2,2). Then, X -< Y but $\pi(X)$ is not majorized by Y. Yet, one can show that if X -< Y and $a_{j+1} \leq b_j$, for all j =1, ..., N-1 then $\pi(X)$ -< Y. The proof is similar to that of Lemma 2.

If we assume that the array Y is decreasing then we have the following refinement of the previous Muirhead-type theorem.

Theorem B: Muirhead for the generalized majorization theorem in a decreasing context

For the N-arrays X and Y in$(\mathbb{R}^+)^N$, X≠Y, and Y decreasing we have:

(i) X -< Y

$\Leftrightarrow$

(ii) Y can be derived from X through a finite number of m (a number depending on X and Y) elementary impact increases:

X = $X^{(0)}$-< $X^{(1)}$-< ... -< $X^{(m)}$, where for each step, j=0, ...,m-2,

$X^{(j)}$ -< $X^{(j+1)}$ -< $\pi(X^{(j+1)})$ -< $X^{(j+2)}$ -< Y and $\pi(X^{(m)}) = Y$ (3)

$\Leftrightarrow$

(iii) Y can be derived from X through a finite number of elementary impact increases

Remark: if m = 1, then of course, the part -< $X^{(j+2)}$ in (3) must be removed.

Proof. (ii) ⇒(iii) is trivial

(iii) ⇒ (i) As X -< EII(X) and as the relation -< is transitive also this implication is trivial.

(i) ⇒ (ii). Let X ≠Y, and

$$X = (x_1, \cdots, x_N) \prec Y = (y_1, \cdots, y_N)$$

In Case A, i.e., if for all $i \in \{1, \ldots, N\}$, $x_i \le y_i$, with, necessarily, at least one index i such that $x_i < y_i$, we proceed as in the case that Y is not necessarily decreasing (using the same notation). Clearly X -< $X^{(1)}$ -< Y and because Y is decreasing: we also have X -< $X^{(1)}$ -< $\pi(X^{(1)})$ -< Y, by Lemmas 1 and 2. Now we can apply the same procedure on $\pi(X^{(1)})$. As the length of the arrays (N) is finite we see that X becomes equal to Y in a finite number of steps and such that in each step the new array is decreasing and situated between the previous one and Y (in the generalized majorization order).

In Case B, i.e., the relation $x_i \le y_i$, for all $i \in \{1, \ldots, N\}$, does not hold, we also start as in the case that Y is not necessarily decreasing, using the same notation. Hence there exists an index j, such that $x_j > y_j$, and $x_j - y_j = c$. After $l$ steps, with $l = 1, \ldots, k$, we have a new array $X_{i_l} = (x_1 + \delta_1, \ldots, x_{i_l} + \delta_{i_l}, x_{i_l+1}, \ldots, x_{j-1}, z_j, x_{j+1}, \ldots, x_N)$. We set $(d_1, \ldots, d_N) = \pi(X_{i_l}) = (x_1 + \delta_1, \ldots, x_{i_l} + \delta_{i_l}, x_{i_l+1}, \ldots, x_{j-1}, x_{j+1}, \ldots, x_{j+a}, z_j, x_{j+a+1}, \ldots, x_N)$, with a ≥ 1. Note that $x_j$ (becoming $z_j$) decreases in each step and hence may move backward in the ranking. However, if $z_j$ does not change ranks, we are in the same situation as in Muirhead's theorem for generalized majorization, and the proof given there can be applied. Now we will show that $\pi(X_{i_l}) \prec Y$.

For s= 1, …, j-1 we have:

$\sum_{k=1}^{s} d_k \le \sum_{k=1}^{s} y_k$ , because, for k = 1, …, i: $d_k = x_k + \delta_k \le y_k$ and for k=i+1, …, j-1: $d_k = x_k \le y_k$ (by the definition of j);

For s = j,…, j+a-1 and for b = 1, …,a:

$\sum_{k=1}^{j+b-1} d_k = \sum_{k=1}^{j-1} x_k + c' + x_{j+1} + \cdots + x_{j+b}$ (recall that c' ≤ $x_j$-$y_j$) ≤

$\left(\sum_{k=1}^{j+b} x_k\right) - y_j \le \left(\sum_{k=1}^{j+b} y_k\right) - y_j \le \left(\sum_{k=1}^{j+b-1} y_k\right)$, where we have used that Y is decreasing and hence $y_j \ge y_{j+b}$ ;

For s = j+a:

$\sum_{k=1}^{j+a} d_k = \sum_{k=1}^{j+a-1} d_k + d_{j+a} = \left(\sum_{k=1}^{j-1} x_k\right) + c' + x_{j+1} + \cdots + x_{j+a} + z_j$ (where we have used the previous equality with b =a)

$$= \left(\sum_{k=1}^{j-1} x_k\right) + c' + x_{j+1} + \cdots + x_{j+a} + x_j - c' = \sum_{k=1}^{j+a} x_k \le \sum_{k=1}^{j+a} y_k$$

and finally, for s ≥ j+a+1:

$$\sum_{k=1}^{s} d_k = \sum_{k=1}^{j+a} d_k + \sum_{k=j+a+1}^{s} x_k = \sum_{k=1}^{j+a} x_k + \sum_{k=j+a+1}^{s} x_k = \sum_{k=1}^{s} x_k \leq \sum_{k=1}^{s} y_k$$

This proves that $\pi(X_{i_l}) - < Y$. Now we have arrived at the same type of inequality as before and we can repeat the same procedure, unless $\pi(X_{i_l}) = Y$ and the theorem is proved. As there are at most a finite number of cases for which $x_j > y_j$, we end up with a proof of the theorem, or with case A, which then can be solved in a finite number of steps. □

An example: X = (4,4,4,4) -< Y = (14,1,1,1). Following the steps of the proof, we have: (4,4,4,4) -< (7,1,4,4) -< (7,4,4,1) -< (10,1,4,1) -< (10,4,1,1) -< (13,1,1,1) -< (14,1,1,1) = Y.

## 4. Conclusion

In this contribution we extended the discrete majorization theory and proved two generalizations of Muirhead's theorem. It is shown how this could play a role in the discrete study of timelines. In the first part, we extend the discrete majorization theory by working with non-normalized Lorenz curves. Then we prove two generalizations of Muirhead's theorem. They not only use elementary transfers but also local increases. Together these operations are described as elementary impact increases. The first generalization shows that if an array X is dominated, in the generalized sense, by an array Y then Y can be derived from X by a finite number of elementary impact increases and this in such a way that each step transforms an array into a new one which is strictly larger in the generalized majorization sense. The other one shows that if the dominating array, Y, is ordered decreasingly then elementary impact increases starting from the dominated array, X, lead to the dominating one. Here each step transforms an array to a new one for which the decreasingly ordered version dominates the previous one and is dominated by Y. It is suggested how this could play a role in the discrete study of timelines.

*Conflict of interest.* The authors declare that they have no conflict of interest.

*Funding.* No funding has been received for this investigation

*Author contributions:*

Leo Egghe: conceptualization, formal analysis, investigation, methodology, writing-original draft, writing-review and editing.
Ronald Rousseau: investigation, validation, writing-review and editing.